\documentclass[12pt]{article}
\usepackage[dvipdfm]{graphicx}
\usepackage{amssymb}
\usepackage{amsmath}
\usepackage{color}
\usepackage{eufrak}

\newcommand{\bea}{\begin{equation}}
\newcommand{\eea}{\end{equation}}
\newcommand{\be}{\begin{eqnarray}}
\newcommand{\ee}{\end{eqnarray}}
\newcommand{\nn}{\nonumber}

\def\hbar#1{\backslash\hspace{-2mm}#1}

\def\nn{\nonumber}

\catcode`\@=11
\def\lsim{\mathrel{\mathpalette\@versim<}}
\def\gsim{\mathrel{\mathpalette\@versim>}}
\def\@versim#1#2{\vcenter{\offinterlineskip
\ialign{$\m@th#1\hfil##\hfil$\crcr#2\crcr\sim\crcr } }}
\catcode`\@=12

\def\2tvec#1#2{
\left(
\begin{array}{c}
#1  \\
#2  \\
\end{array}
\right)}

\def\mat2#1#2#3#4{
\left(
\begin{array}{cc}
#1 & #2 \\
#3 & #4 \\
\end{array}
\right) }

\def\Mat3#1#2#3#4#5#6#7#8#9{
\left(
\begin{array}{ccc}
#1 & #2 & #3 \\
#4 & #5 & #6 \\
#7 & #8 & #9 \\
\end{array}
\right) }

\def\3tvec#1#2#3{
\left(
\begin{array}{c}
#1  \\
#2  \\
#3  \\
\end{array}
\right)}

\def\hbar#1{\backslash\hspace{-2mm}#1}

\numberwithin{equation}{section}

\begin{document}

\begin{titlepage}
\begin{flushright}
KANAZAWA-12-02, P12019
\end{flushright}

\begin{center}

\vspace{1cm}
{\large\bf Can A Higgs Portal Dark Matter be Compatible with the
 Anti-proton Cosmic-ray?}
\vspace{1cm}

Hiroshi Okada$^{a}$\footnote{HOkada@kias.re.kr}
 and
Takashi Toma$^{b,c}$\footnote{t-toma@hep.s.kanazawa-u.ac.jp}
\vspace{5mm}

{\it%
$^{a}${School of Physics, KIAS, Seoul 130-722, Korea}\\
$^{b}${ Institute for Theoretical Physics, Kanazawa University, Kanazawa, 920-1192, Japan}\\
  $^{c}${Max-Planck-Institut f\"ur Kernphysik, Saupfercheckweg 1, 69117
 Heidelberg, Germany}}
  
  \vspace{8mm}

\abstract{
Recent direct detection experiments of Dark Matter (DM), CoGeNT and DAMA
 implicate a light DM of a few GeV. Such a light DM would generate a
 large amount of anti-proton since suppression for anti-proton flux from DM
 annihilation is ineffective.
We discuss whether a light dark matter with mass of 
$5-15~{\rm GeV}$, which is especially in favor of the recent experiments reported
 by CoGeNT, is compatible with the anti-proton no excess
 in the cosmic-ray. 
In view of the direct detection of DM and no anti-proton excess in the
 cosmic-ray both, we show that a Dirac DM is favored than a scalar one
 since there is no s-wave of the annihilation cross section for the
 Dirac DM. A large elastic cross section for direct detection can be
 obtained through the additional light Higgs exchange. We show an
 allowed region that simultaneously satisfies the
 DM relic density, the elastic cross section favored by CoGeNT and also
 the constraint of $H_LZZ$ coupling of the light Higgs boson by LEP. 

 }

\end{center}
\end{titlepage}

\setcounter{footnote}{0}

\section{Introduction}
The existence of Dark Matter (DM) is crucial from cosmological
observations such as the rotation curves of the
galaxy~\cite{Begeman:1991iy}, the CMB observation by WMAP~\cite{wmap}, gravitational
lensing~\cite{Massey:2007wb}. However, since DM candidate particle is not
included in the Standard Model (SM), the SM should be extended to
include DM candidate. 
In recent years, direct detection experiments of
DM such as XENON100~\cite{Aprile:2011hi}, CDMSII~\cite{Ahmed:2009zw},
CRESSTII~\cite{Angloher:2011uu}, CoGeNT~\cite{cogent} and
DAMA~\cite{Bernabei:2010mq} are extremely active to 
look for scattering events with nuclei. While XENON100 and CDMSII have shown
null result of DM signal, CoGeNT, DAMA and CRESSTII have reported 
the observations which can be interpreted as DM signals. 
The discrepancy among these experiments could be improved by considering
a non-standard DM scattering model like inelastic
scattering~\cite{TuckerSmith:2001hy, TuckerSmith:2004jv}, or
a non-spherical DM profile like tri-axiality~\cite{Gustafsson:2006gr,
Ling:2009eh}. 
The CoGeNT and DAMA results observing annual modulation of
DM signals favor a light DM with several GeV mass and rather large
scattering cross section with nuclei.  
If the CoGeNT and DAMA results are truly DM signals, we need a DM model
which is consistent with the other constraints such as the DM
thermal relic density, the observation of anti-proton cosmic-ray. 
In case of taking into account a light DM, 
anti-proton emission in the cosmic-ray due to annihilation of
DM would increase. This is because
the source term of the anti-proton flux from DM annihilation is
$\left<\sigma v_{\mathrm{rel}}\right>\rho_{\odot}^2/2m_{DM}^2$, where
$\left<\sigma v_{\mathrm{rel}}\right>$ is annihilation cross section of
DM, $\rho_{\odot}\simeq0.3~\mathrm{GeV/cm^3}$ is the local DM density
at the Earth and $m_{DM}$ is DM mass. Thus we can see that suppression of
the source term by DM mass does not work efficiently for
$10~\mathrm{GeV}$ scale DM. 

The minimal extension of the SM which includes DM candidate is the model
with Higgs portal gauge singlet real scalar
$S$~\cite{Andreas:2010dz}. This model is fascinating since the annihilation cross section
required to make the correct DM relic density and the elastic cross section
with nuclei favored by CoGeNT and DAMA are simultaneously obtained at roughly
$10~\mathrm{GeV}$ DM mass. However if the constraint from the
anti-proton cosmic-ray should be taken into account, this model
is severely restricted~\cite{Evoli:2011id, arXiv:1007.5253, Kappl:2011jw}. 
In order to escape the anti-proton constraint, the annihilation cross
section of DM should be p-wave dominant. If so, the annihilation cross
section is extremely suppressed by the relative velocity
$v_{\mathrm{rel}}\sim 10^{-3}$ at the present Universe. In this paper, we discuss a
next minimal DM model which has a gauge singlet real
scalar $S$ and a Dirac fermion $\Psi$, and the Dirac fermion is identified
to be DM candidate with p-wave dominant annihilation cross
section\footnote{Other scenarios are considered overcome it. See,
e.g., Ref.~\cite{Goudelis:2009zz, Keung:2010tu, Cerdeno:2011tf}}. 
In this scenario, the singlet real scalar $S$ is assigned as
even charge under the $\mathbb{Z}_2$ parity which is different from the
minimal DM model. Thus the SM model Higgs $h$ and the singlet real
scalar $S$ mix and the mixing is limited by the $H_LZZ$ coupling
constraint from the LEP experiment~\cite{lep} where $H_L$ is light mass
eigenstate of Higgs. 
The elastic scattering of DM with nuclei for direct detection occurs
through the Higgs exchange, hence the contribution through the light
Higgs becomes larger than that through the heavy Higgs. As a result, a large
elastic cross section of DM required from CoGeNT is expected to be obtained.
We investigate whether the DM relic density and
the rather large elastic cross section favored by CoGeNT are satisfied
simultaneously in the allowed parameter space from the Higgs mixing
bound by LEP.

This paper is organized as follows. In Section 2, we review the minimal
Higgs portal DM model and discuss the next minimal DM model with a Dirac
fermion. 
In Section 3, direct detection of DM in the next minimal DM
model is discussed. We summarize and conclude in Section 4. 


\section{The Model}

\subsection{The Minimal Model}
At first we start to discuss a gauge singlet real scalar DM ($S$) in the simplest model.
The Lagrangian is typically given as
\be
{\cal L}\!\!\!&=& y_u \overline{Q} \tilde H U_R+y_d \overline{Q} H D_R +
y_\ell \overline{L} H E_R + \mathrm{h.c.}\nonumber\\
&&\!\!\!-\frac{m^2_{S}}{2}S^2 - m^2_HH^\dagger H -\frac{\lambda_H}{2}(H^\dagger H)^2 - \frac{\lambda_S}{4}S^4
-\frac{\lambda_{SH}}{2} S^2 \left(H^\dagger H\right),
\ee 
where the odd $\mathbb{Z}_2$ parity is imposed to $S$ to guarantee the
DM stability and the family index is
omitted. Here $H$ is SM Higgs and $\tilde H\equiv i\sigma_2 H^*$. 
Note that $S$ is the mass eigenstate at this Lagrangian since it does
not mix with SM Higgs because of the $\mathbb{Z}_2$ parity.
After the spontaneous symmetry breaking; $H=(0,v+h/\sqrt2)^t$ with $v$=174 GeV,
one finds the interacting terms between Higgs and matter fields as 
\begin{equation}
{\cal L}\ni  \frac{y_u }{\sqrt2}\overline{U_L} U_Rh+\frac{y_d
 }{\sqrt2}\overline{D_L} D_Rh + \frac{y_\ell}{\sqrt2} \overline{E_L} E_R h
- \frac{\lambda_{SH} v}{\sqrt2} S^2 h +\mathrm{h.c.}.
\end{equation}
The singlet scalar $S$ annihilates to the SM fermions $f\overline{f}$
through the neutral Higgs boson exchange.
 Here notice that 
the dominant contribution comes from bottom pair final state as one can
 see from the hierarchy of the Yukawa coupling, thus we consider the
 $b\overline{b}$ process only hereafter. 
The annihilation cross section of the singlet scalar is calculated as
\begin{eqnarray}
\sigma v_{\mathrm{rel}}
\!\!\!&\simeq&\!\!\!
\frac{3\lambda_{SH}^2 v^2}{16\pi}\left(\frac{1}{s-m^2_h}\right)^2 y^{2}_b 
\left(1-\frac{m^2_b}{m^2_{DM}}\right)
\left[\left(1-\frac{m^2_b}{m^2_{DM}}\right)+\left(\frac{3m^2_b}{8m^2_{DM}}\right)v^2_{\mathrm{rel}}\right]
+\mathcal{O}(v_{\mathrm{rel}}^4),
\label{eq:ann0}
\end{eqnarray}
 where $m^2_{DM}=m^2_S+\lambda_{SH}v^2$ and 
 $v_{\mathrm{rel}}$ is the relative velocity of the annihilating DM. 

\begin{figure}[t]
\begin{center}
\includegraphics[scale=0.43]{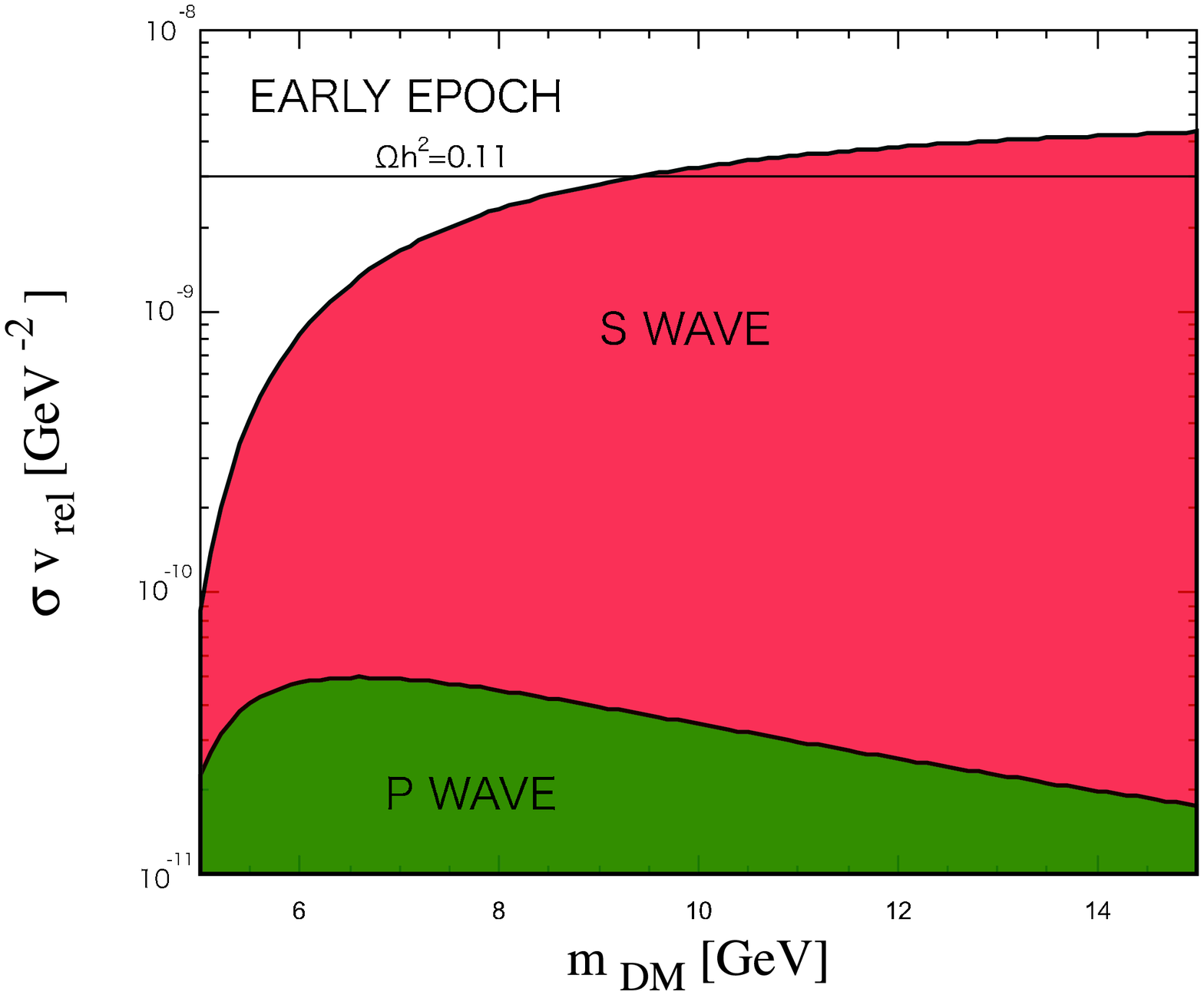}
\includegraphics[scale=0.43]{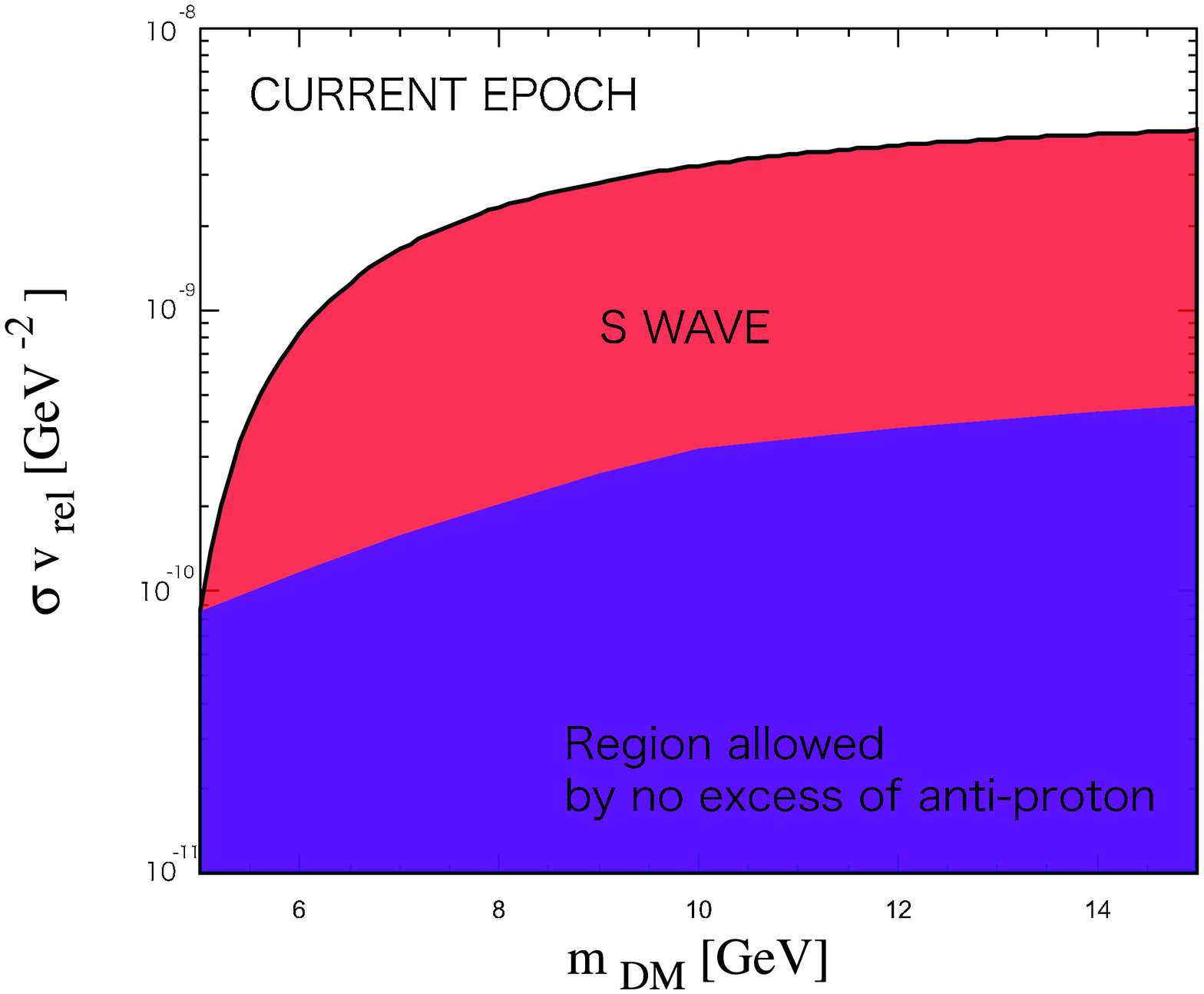}
\caption{
The left figure is s-wave and p-wave contribution of the annihilation cross
 section Eq.~(\ref{eq:ann0}) in the early Universe epoch. 
The horizontal black line is the annihilation cross section;
 $\sigma v_{\mathrm{rel}}= 3\times10^{-9}~\mathrm{GeV}^{-2}$, satisfying
 the DM relic density $\Omega h^2=0.11$ from the WMAP~\cite{wmap}. 
 The right figure is the s-wave contribution and the allowed
 parameter region from the anti-proton no excess in the cosmic-ray
 observations. Notice that the p-wave contribution is negligible
 compared with the s-wave one. The blue region is
 allowed from the anti-proton constraint in the case of the NFW DM
 profile~\cite{Navarro:1996gj}.}
\label{sp-wave}
\end{center}
\end{figure}

The left panel of Fig.~\ref{sp-wave} shows s-wave and p-wave
 contribution of Eq.~(\ref{eq:ann0}) at the early Universe
 ($v_{\mathrm{rel}}\sim 0.3$) where we take $m_b=4.67$ GeV~\cite{PDG2010},
 $\lambda_{SH}\le1$ and $m_h=125~\mathrm{GeV}$~\cite{cms-atlas}.
The horizontal black line implies the annihilation cross section required to get
 the correct DM relic density from WMAP, which is roughly $\sigma
 v_{\mathrm{rel}}\sim3\times10^{-9}~\mathrm{GeV}^{-2}$. 
 In the right panel of Fig.~\ref{sp-wave}, the s-wave contribution at
 the present Universe and allowed parameter region from the
 anti-proton no excess in the cosmic-ray
 observations~\cite{Adriani:2010rc, Aguilar:2002ad, Asaoka:2001fv,
 Maeno:2000qx, Orito:1999re} are shown. The p-wave contribution can be
 neglected due to $v_{\mathrm{rel}}\sim 10^{-3}$ at the time.
As shown in the figure, the s-wave contribution should be less than
 ${\cal O}$ (10$^{-10}$) GeV$^{-2}$ to satisfy the
anti-proton no excess if a few GeV DM is taken into account. 
From these figures, we can see that the annihilation
cross section in the minimal singlet scalar model cannot satisfy both
 of the relic density required by WMAP and the anti-proton no excess in
 the cosmic-ray at the present Universe. 
It suggests that the annihilation cross section required by WMAP
 should be generated by p-wave contribution. 
Hence we conclude that the minimal model cannot be consistent with the
 DM relic density of the light DM and no anti-proton excess in the
 cosmic-ray. 
 Note that there is an ambiguity for the anti-proton
 constraint which depends on DM profiles such as
 NFW, Einasto and Burkert and on propagation models of cosmic-ray.
 In particular the dependence on the vertical scale of the diffusion zone
 is significant. If we choose an only small value of the parameter, it is
 possible to keep consistency with the constraint. 
 This is discussed in \cite{Evoli:2011id} in detail.

The minimal model also strongly affects the SM Higgs invisible
decay~\cite{Burgess:2000yq, Barger:2007im, Andreas:2008xy, He:2009yd}. 
The large $hSS$ coupling gives an impact to the branching ratio of the
Higgs invisible decay and can be dominant Higgs decay channel when the
mass of the singlet scalar $S$ is less than
$40~\mathrm{GeV}$~\cite{Burgess:2000yq}. It makes difficult to search an
evidence of DM in the minimal model by collider experiments.


\subsection{The Next Minimal Model}
In order to solve the problem of the anti-proton cosmic-ray, 
we propose a next minimal model, in which
only the difference is that a Dirac fermion $\Psi$ is newly added to the
minimal singlet scalar model where the $\mathbb{Z}_2$ parity is imposed
to odd for the $\Psi$ and even for the other particles including the
singlet scalar $S$. Thus the Dirac fermion $\Psi$ can be a DM candidate in
this model. 
It is worth mentioning that in the same model S. Baek, P. Ko, and Wan-Il
Park group have recently discussed to analyze the other aspects such as
perturbative unitarity, electroweak precision test and LHC detectability
with direct DM search at ${\cal O}$(100) GeV, in which they have shown
an good agreement with those experiments \cite{Baek:2011aa}. 
The singlet scalar $S$ cannot be a DM in the
next minimal model due to the even assignment of the $\mathbb{Z}_2$ parity. 
The additive Lagrangian to the SM is written as
\begin{eqnarray}
\mathcal{L}\!\!\!&=&
 gS\overline{\Psi}\Psi - m_{DM}\overline{\Psi}\Psi - m_H^2H^\dag H
 - \frac{m_S^2}{2}S^2
 -\lambda_H\left(H^\dag H\right)^2\nonumber\\
 &&\!\!\!
-\mu_SS(H^\dagger H) - \frac{\lambda_{SH}}{2}
 S^2\left(H^\dag H\right)
- \mu_1^3 S - \frac{\mu_2}{3} S^3-\frac{\lambda_S}{4}
  S^4. 
\end{eqnarray}
Here notice that Higgs sector is no longer the mass eigenstate 
because Higgs mixing term appears from the interaction
$SH^{\dag}H$~\footnote{
Although in general the pseudo scalar coupling such as
$S\overline{\Psi}\gamma_5\Psi$ also exists, here
we do not consider it since the pseudo scalar coupling generates s-wave
to the DM annihilation cross section.
}.
Assuming all the parameters in the Higgs sector are real, that is, there
is no CP violation, we find the Higgs mass matrix in the basis of $(h,S)^t$. It can be
diagonalized by the following mixing matrix as
\begin{eqnarray}
 M^2_{\rm Higgs} &\equiv&
 \left(\begin{array}{cc} 4\lambda_Hv^2 & 2\sqrt{2}\mu_Sv \\
       2\sqrt{2}\mu_Sv  & m^2_S+2\lambda_{SH} v^2 \end{array}\right)\nonumber\\
 &=& \left(\begin{array}{cc} \cos\alpha & \sin\alpha \\ -\sin\alpha & \cos\alpha \end{array}\right)
\left(\begin{array}{cc} m^2_{H_{L}} & 0 \\ 0 & m^2_{H_{H}}  \end{array}\right)
\left(\begin{array}{cc} \cos\alpha & -\sin\alpha \\ \sin\alpha &
      \cos\alpha \end{array}\right),
\label{eq:higgs}
\end{eqnarray}
where we inserted the stable condition: $m^2_H = -2\lambda_H v^2$, $\mu_1^3=-\mu_Sv^2$ and 
we assumed that $S$ has no vacuum expectation value.
The gauge eigenstates $h$ and $S$ can be rewritten in terms of the mass eigenstates of $H_L$ and $H_H$
as
\begin{eqnarray}
 h &=& H_{L} \cos\alpha + H_{H}\sin\alpha, \nn\\
 S &=&- H_{L} \sin\alpha + H_{H} \cos\alpha.
\label{eq:mass_weak}
\end{eqnarray}
Then the Lagrangian includes the interactions with the light Higgs boson $H_L$
\begin{equation}
{\cal L}=
 -g\sin\alpha \overline{\Psi}\Psi H_{L}+
\frac{y_b \cos\alpha}{\sqrt2}\overline{b_L} b_R H_{L}
+ \mathrm{h.c.},
\end{equation}
where we omit the heavy Higgs contribution. 
From the interactions, the annihilation cross section for
$\overline{\Psi}\Psi \to \overline{f}f$ through the light scalar boson
$H_L$ can be written as
\begin{equation}
\sigma v_{\mathrm{rel}}\simeq \frac{3g^2y^2_b m_{DM}^2}{16 \pi} 
\frac{\sin^2\alpha\cos^2\alpha}{\left(s-m_{H_L}^2\right)^2+m_{H_L}^2\Gamma_{H_L}^2}
 \left(1-\frac{m_b^2}{m_{DM}^2}\right)^{3/2} v^2_{\mathrm{rel}}+\mathcal{O}(v^4_{\mathrm{rel}}).
\label{eq:ann}
\end{equation}
where $\Gamma_{H_L}$ is the total decay width of the light Higgs
$H_L$. When $2m_{b}<m_{H_L}<2m_{DM}$, the decay
width is 
\begin{equation}
\Gamma_{H_L}\simeq
\frac{y_b^2\cos^2\alpha}{8\pi}m_{H_L}\left(1-\frac{4m_b^2}{m_{H_L}^2}\right)^{3/2},
\end{equation}
and when $2m_{DM}<m_{H_L}$, the channel $H_L\to\mathrm{DM\:DM}$ is also added.
we neglected the light quarks contributions and put only bottom quark contribution. 
If DM is heavier than the light Higgs $H_L$, the annihilation
channel $\mathrm{DM\:DM}\to H_LH_L$ is also possible and it is large
enough to obtain proper relic abundance of DM. However as we will
see Fig.~\ref{fig:wmap-dd} in Section~\ref{secdirect}, since the light Higgs mass should be
heavier than DM in order to fit the CoGeNT result, we do not take
into account the channel here. 
The annihilation to $b\overline{b}$ is dominant as same as the minimal
singlet scalar model. 
We can obtain no s-wave in the annihilation cross section, thus 
the anti-proton constraint from the cosmic-ray at the present Universe
is consistent as we discussed before.

\begin{figure}[t]
\begin{center}
\includegraphics[scale=0.80]{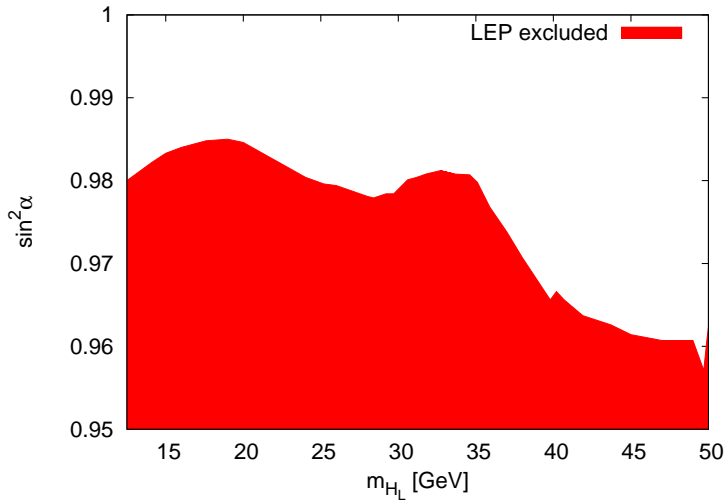}
\caption{LEP bound of the light neutral Higgs boson $H_L$. 
The red region is excluded by the LEP data.}   
\label{lep}
\end{center}
\end{figure}

The Higgs mixing $\sin\alpha$ is strongly constrained by the $H_LZZ$
coupling measurement. The LEP data are used to set upper bound on the $H_LZZ$
coupling~\cite{lep}. The excluded region of $\sin^2\alpha$ by LEP data
in $10<m_{H_L}<50~\mathrm{GeV}$ is shown in Fig.~\ref{lep}. 
The light Higgs mass should be $m_{H_L}\lesssim50~\mathrm{GeV}$ to be
consistent with the required annihilation cross section by WMAP and the
elastic cross section favored from CoGeNT as discussed later. 
Thus the mass range of Fig.~\ref{lep} is enough in the discussion. 
We can see from the Fig.~\ref{lep} that region of $\sin\alpha\simeq1$ is
only allowed from the LEP bound in this mas region. This implies that
the off-diagonal element of the Higgs mass matrix Eq.~(\ref{eq:higgs})
must be small, namely small $\mu_S$.

The light Higgs $H_L$ might affect to the decay of the heavy Higgs boson
which corresponds to the SM Higgs boson. 
The new invisible decay channel $H_H\to H_LH_L$ appears and 
the branching ratio of the channel is proportional to the coupling
$\lambda_{SH}\sin^3\alpha$. If the branching fraction of the invisible
decay $H_H\to H_LH_L$ is too large, the next minimal model is ruled out
by the recent LHC result as analyzed in the other
models~\cite{AlbornozVasquez:2011aa}. 
However fortunately, we can take a small
value of the coupling since the coupling $\lambda_{SH}$ is not relative
with DM physics which is different from the minimal model. 

The cusp profile such as NFW is favored from the N-body simulations of
collisionless DM. On the other hand, the X-ray observations of clusters 
suggest that DM density profile is approximately flat. 
The discrepancy between the simulations and the observations of the
galaxies might be alleviated by taking into account the self-interaction
of DM~\cite{Spergel:1999mh, Markevitch:2003at, Taoso:2007qk,
Loeb:2010gj}. 
The elastic cross section of DM must satisfy
$\sigma/m_{DM}\lesssim 1~\mathrm{cm^2/g}$ to be consistent with
the observations. The constraint becomes
important in the case of considering $\mathrm{MeV}$ scale DM or some
enhancement mechanism of elastic (annihilation) cross section of DM like
Sommerfeld enhancement~\cite{Buckley:2009in}. 
In the model, t-channel through the light Higgs $H_L$ is dominant for
the elastic cross section of DM, and it is rather enhanced by
intermediating the light Higgs. However the enhancement is not so large
and we do not need to take into account the constraint from the
self-interaction of DM. 

\section{Direct Detection}\label{secdirect}
\begin{figure}[t]
\begin{center}
\includegraphics[scale=1.0]{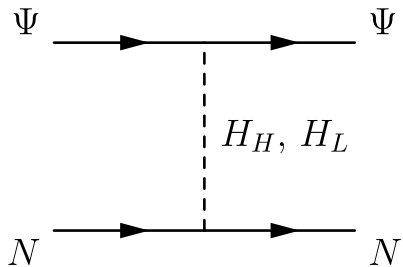}
\caption{The t-channel diagram for the direct detection of DM.}   
\label{dd-diag}
\end{center}
\end{figure}

We analyze the direct detection search of DM through CoGeNT
\cite{cogent}. The global fit analysis in DM mass and elastic
cross section plane has been performed in Ref.~\cite{Hooper:2012ft}. 
In this model the main contribution to the spin-independent (SI) cross
section comes from the t-channel diagram 
intermediated by the light Higgs $H_L$ as depicted in Fig.~\ref{dd-diag}. 
Then the resultant SI elastic cross section with nucleon is
given by 
\begin{equation}
\sigma_{\mathrm{SI}}^N\simeq\frac{\mu_{\mathrm{DM}}^2}{\pi}
\left(\frac{gm_N\sin\alpha\cos\alpha}
{m_{H_L}^2v}\sum_{q}f_q^N\right)^2,
\label{eq:dd}
\end{equation}
where $\mu_{\mathrm{DM}}=\left(m_{DM}^{-1}+m_N^{-1}\right)^{-1}$ is the
DM-nucleon reduced mass and the heavy Higgs contribution is neglected. 
We can see from the Eq.~(\ref{eq:dd}) that the elastic cross section
vanishes if $\sin\alpha$ approaches to $1$ because of $\cos\alpha\to 0$. 
This behavior is same as that for the annihilation cross section
Eq.~(\ref{eq:ann}). We can obtain a large SI elastic cross section due to the
propagation of the light Higgs. The parameter $f_q^N$ stands for the
contribution of each quark to nucleon mass and these are calculated
from the lattice simulation~\cite{Ohki:2008ff, Corsetti:2000yq} as
\begin{eqnarray}
&&f_u^p=0.023,\quad
f_d^p=0.032,\quad
f_s^p=0.020,\\
&&f_u^n=0.017,\quad
f_d^n=0.041,\quad
f_s^n=0.020,
\end{eqnarray}
for the light quarks. For the heavy quarks, the parameters are given as
$f_Q^N=2\left(1-\sum_{q\leq 3}f_q^N\right)/27$ where $Q$ stands for the
heavy quarks and $q\leq3$ implies the summation of the light quarks.

\begin{figure}[t!]
\begin{center}
\includegraphics[scale=0.35]{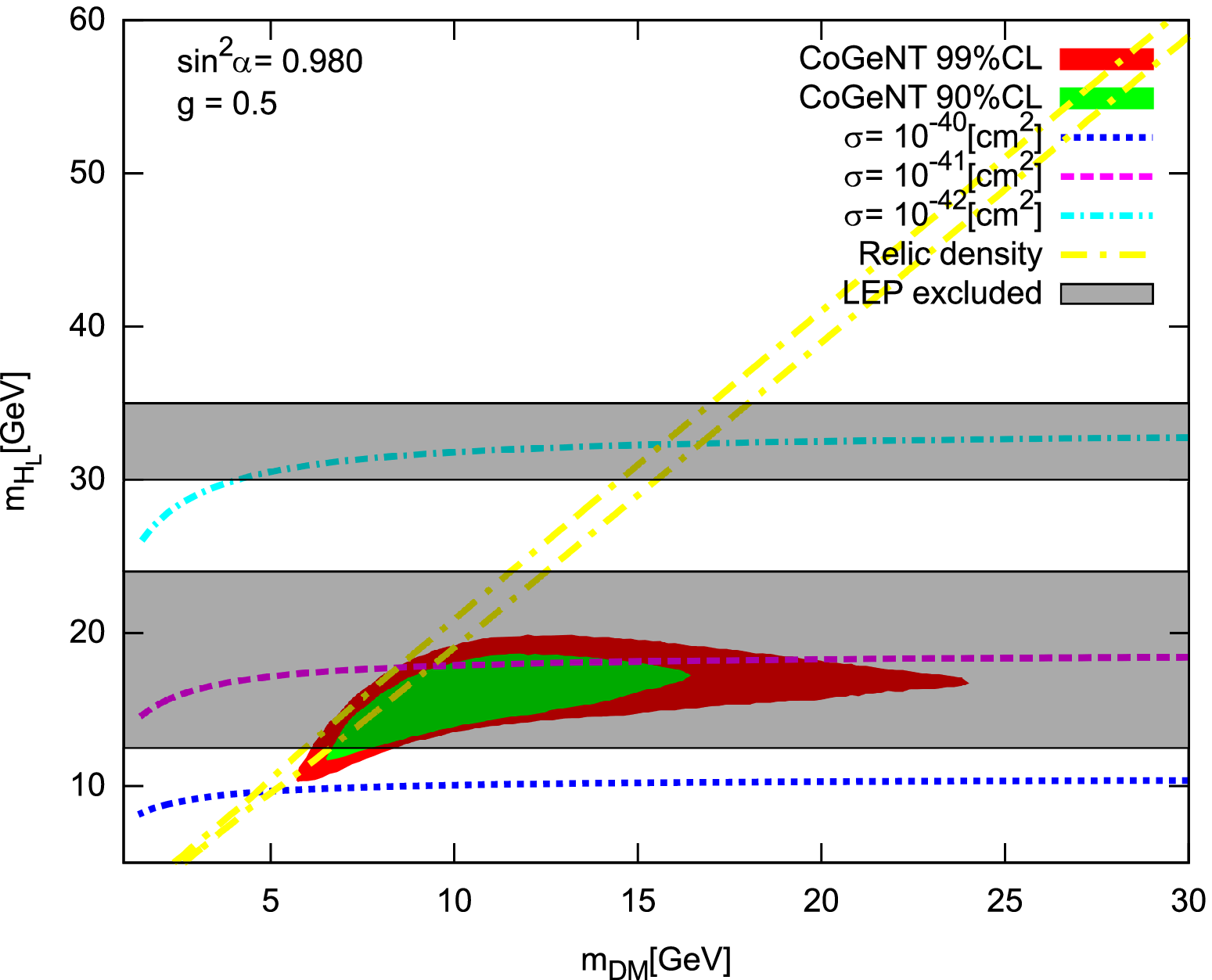}
\quad
\includegraphics[scale=0.35]{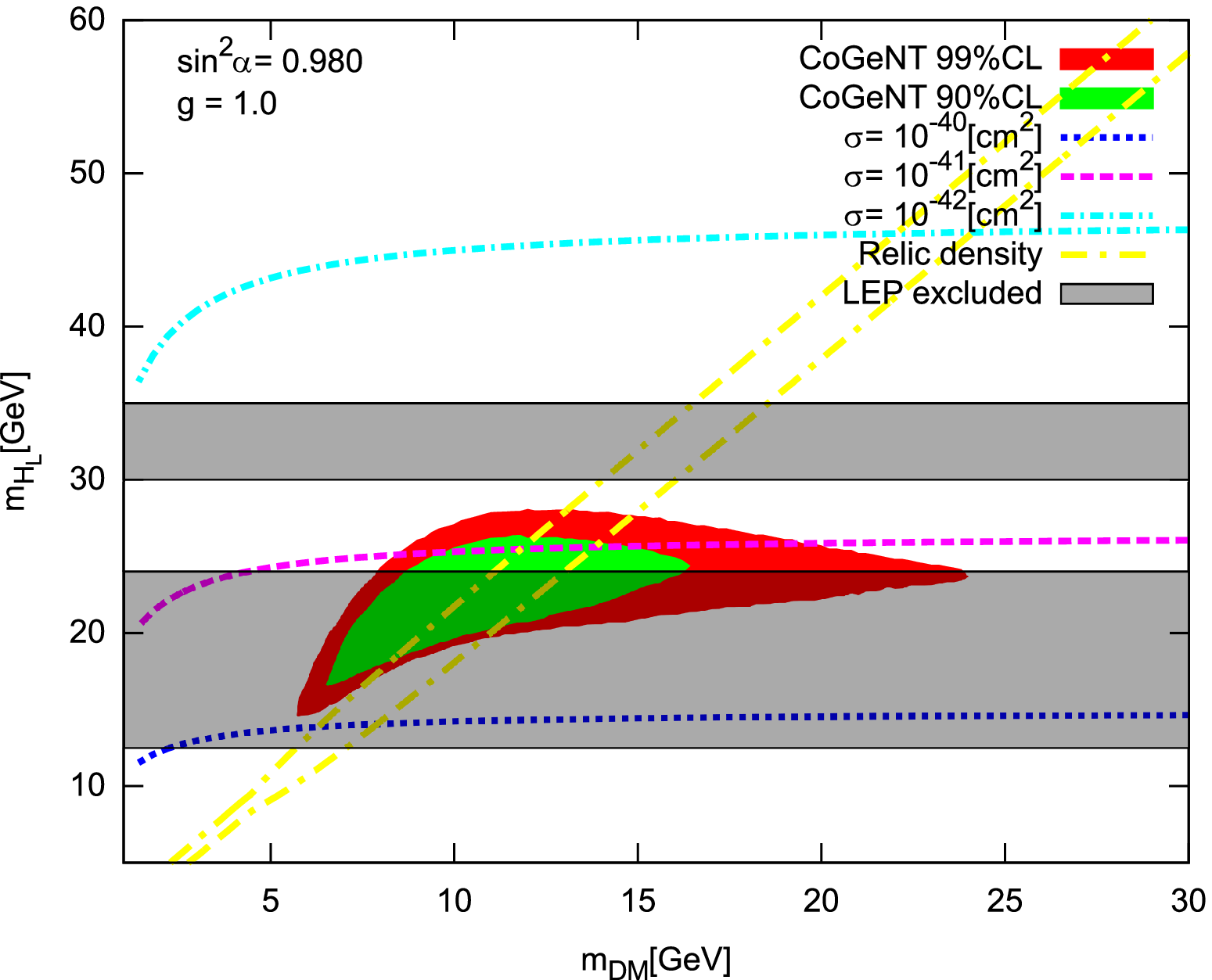}
\quad
\includegraphics[scale=0.35]{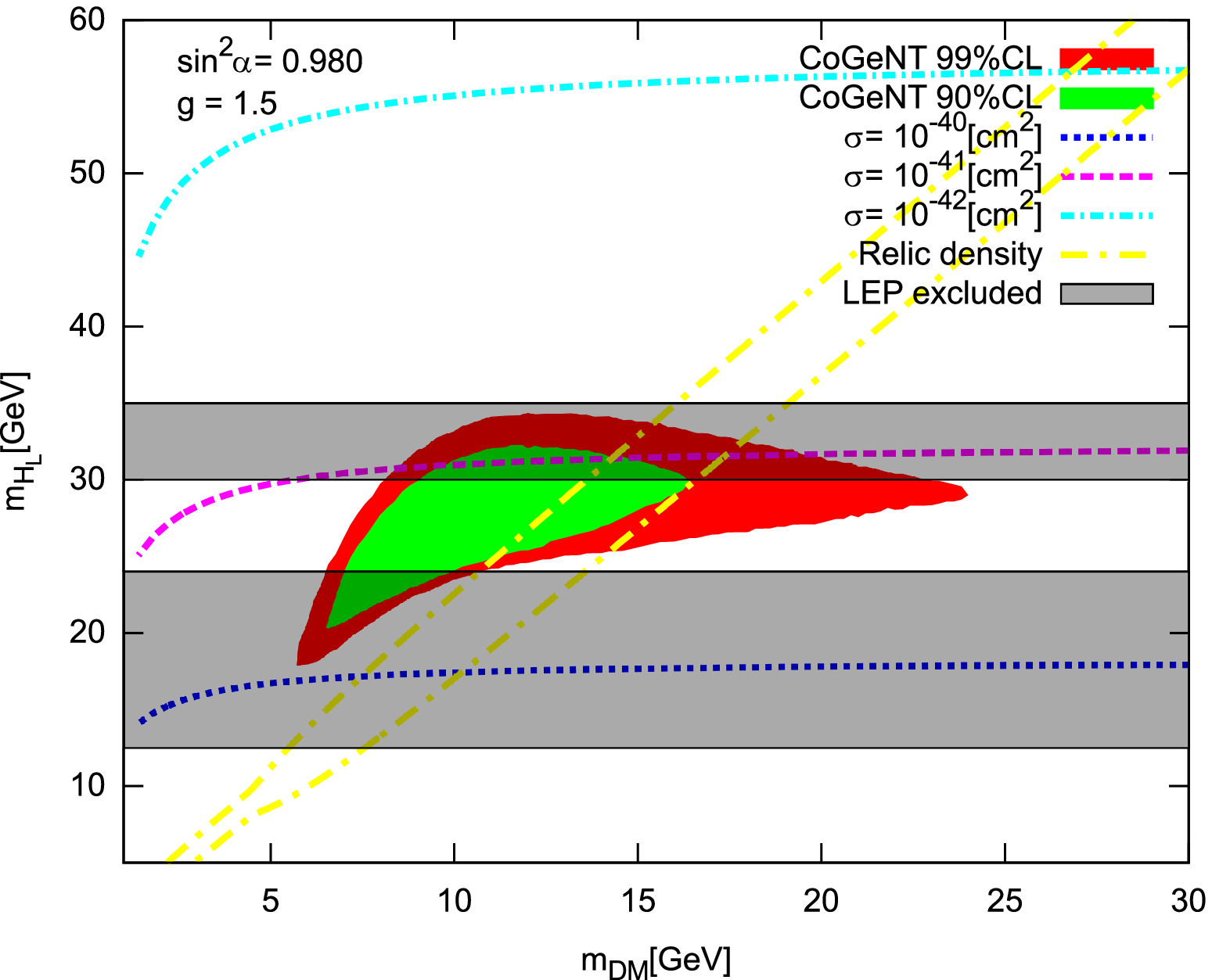}

\vspace{0.5cm}
\includegraphics[scale=0.35]{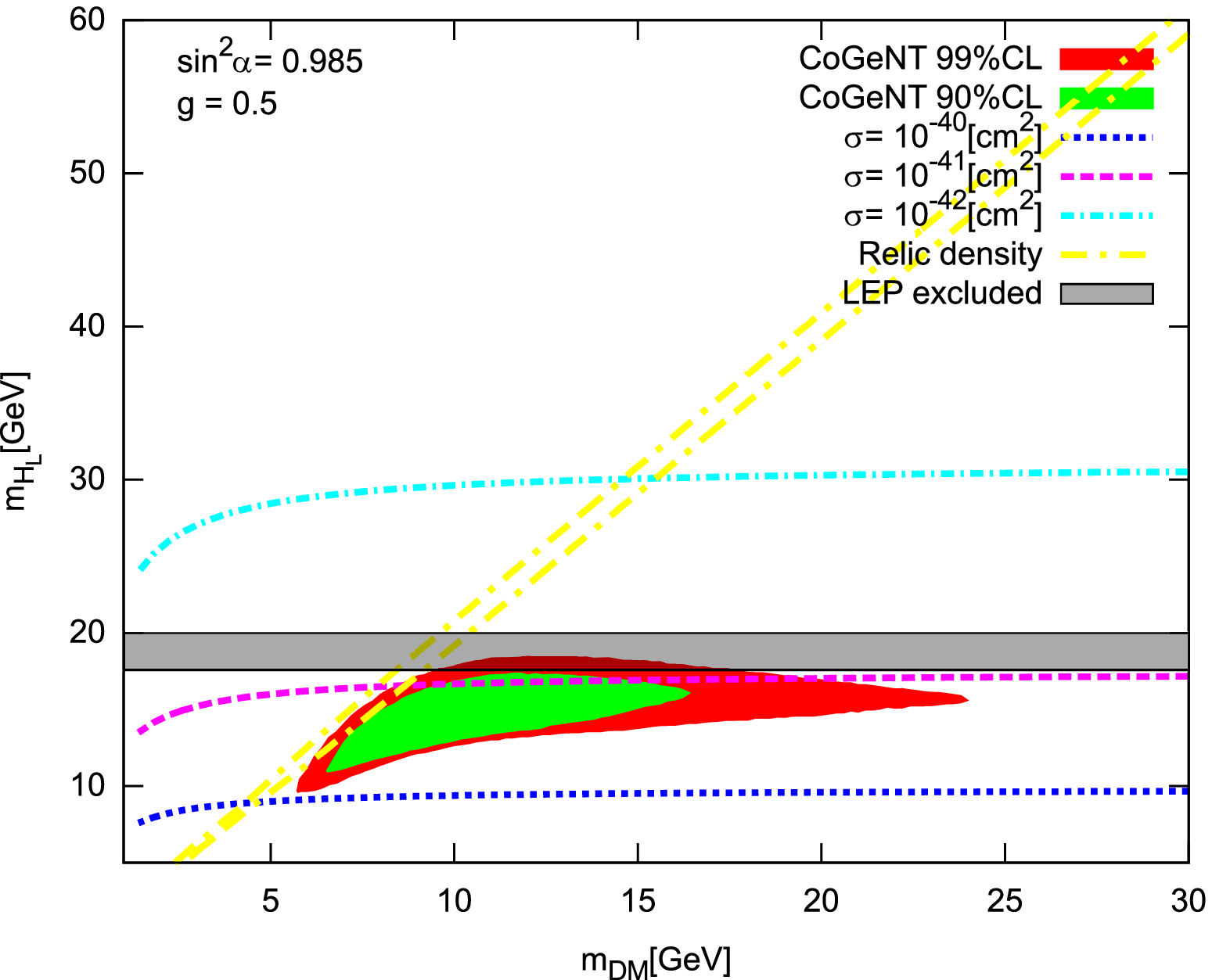}
\quad
\includegraphics[scale=0.35]{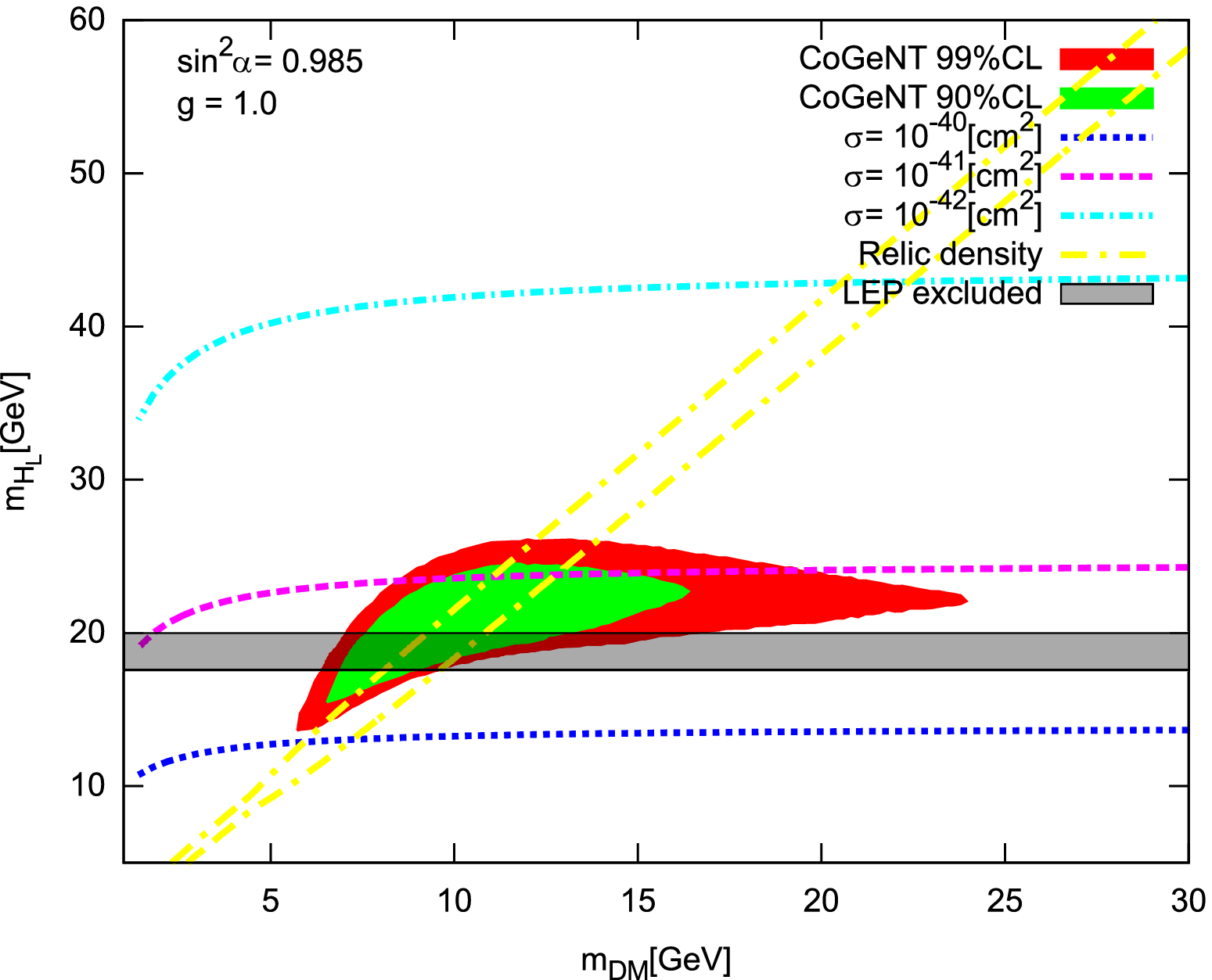}
\quad
\includegraphics[scale=0.35]{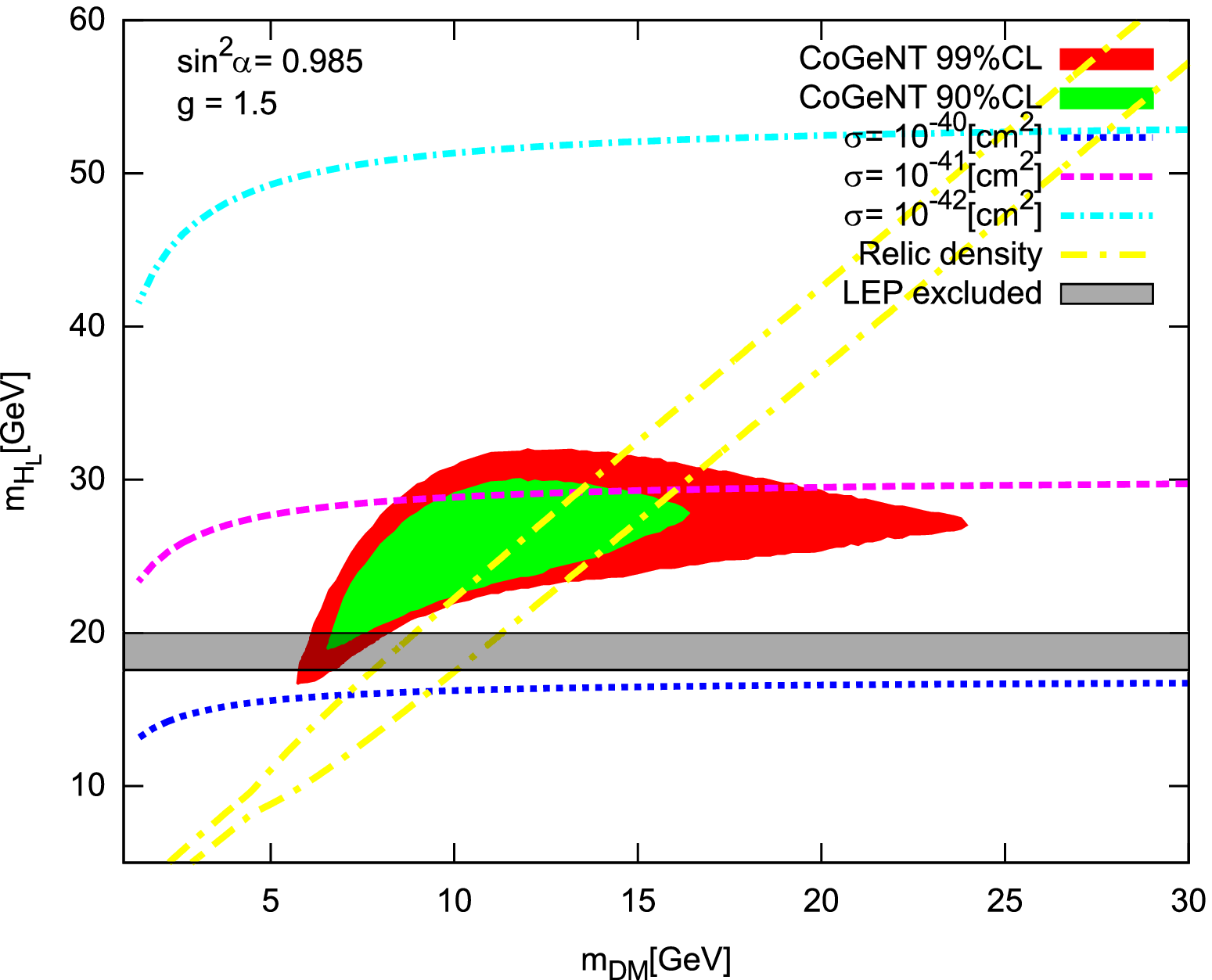}

\vspace{0.5cm}
\includegraphics[scale=0.35]{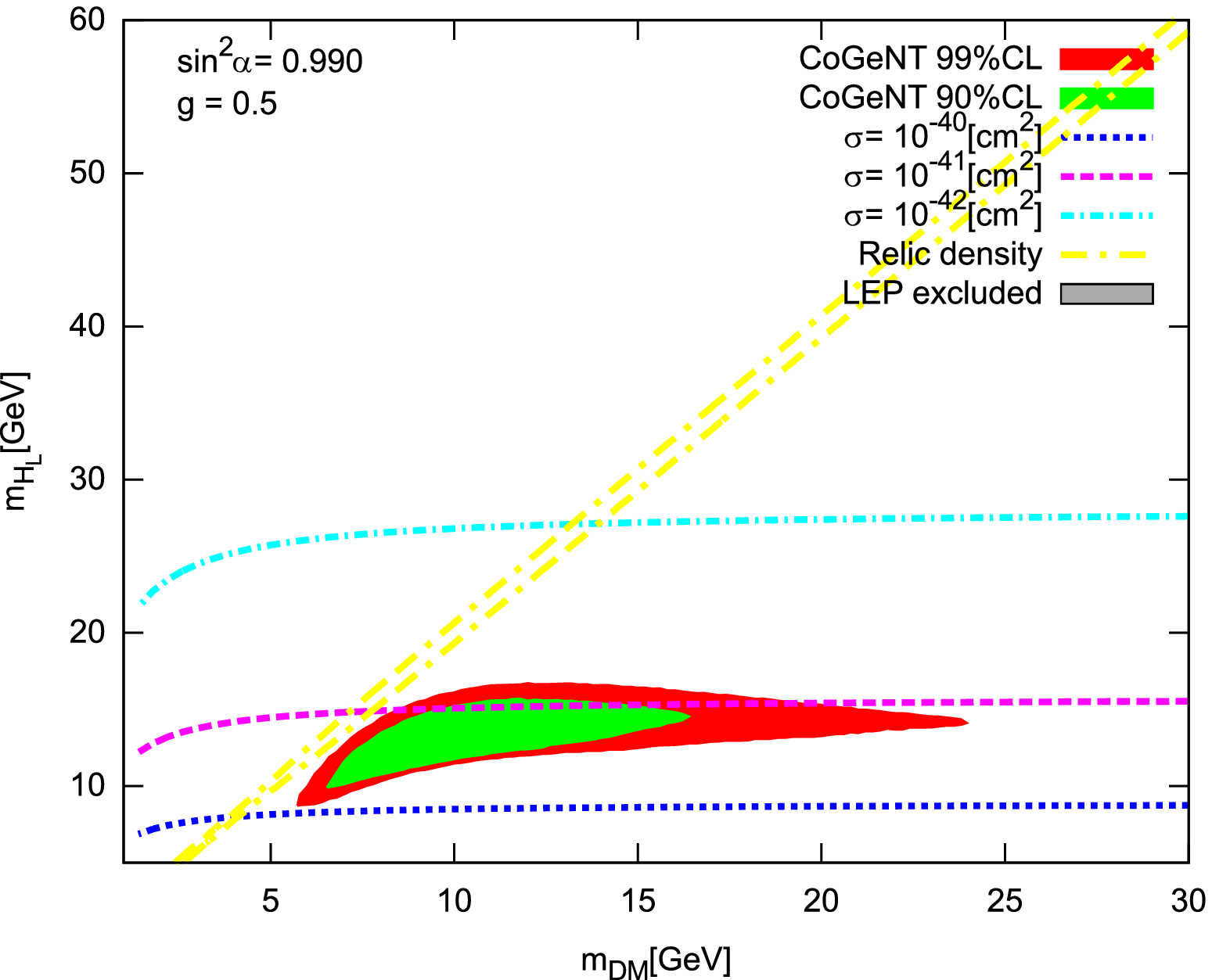}
\quad
\includegraphics[scale=0.35]{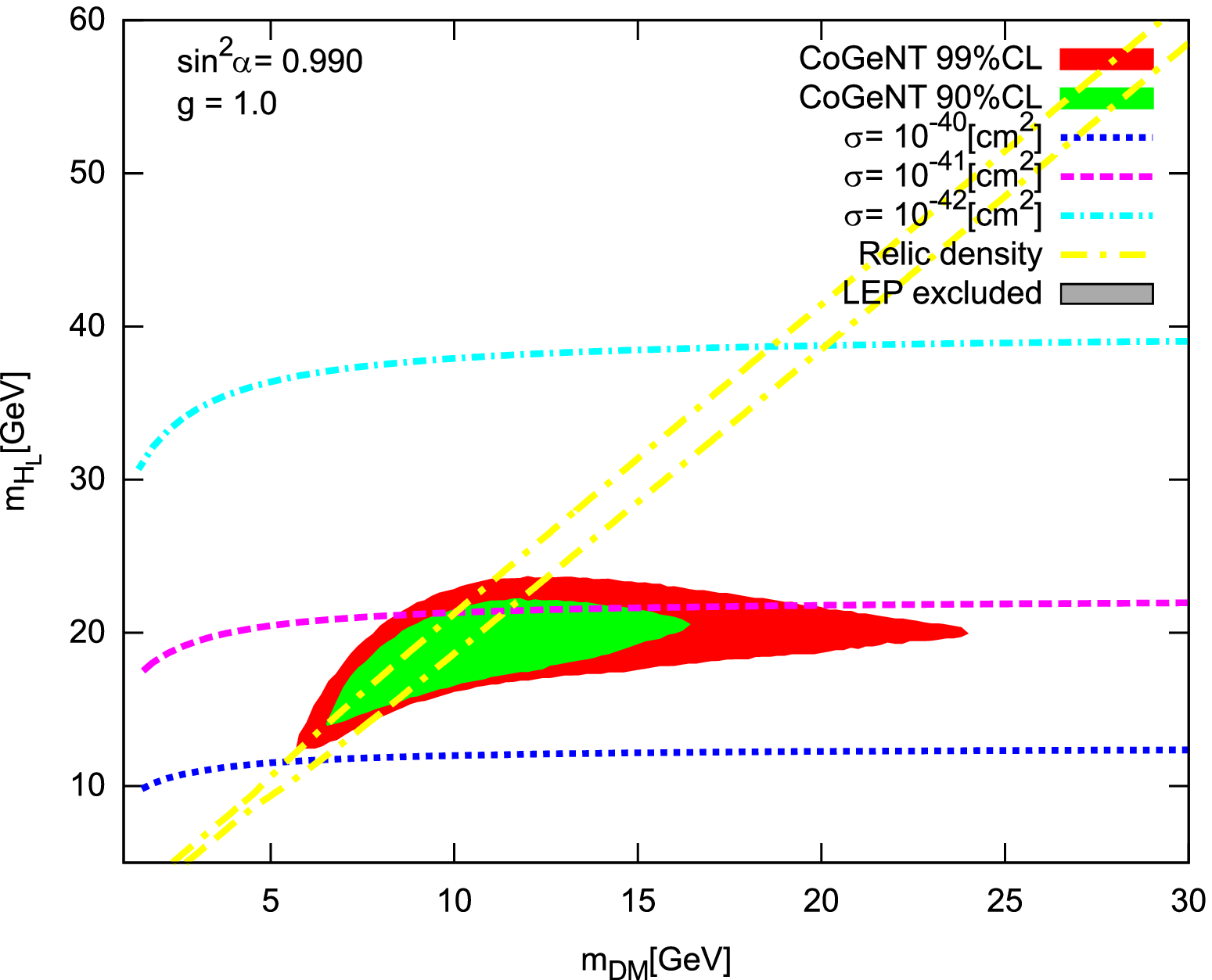}
\quad
\includegraphics[scale=0.35]{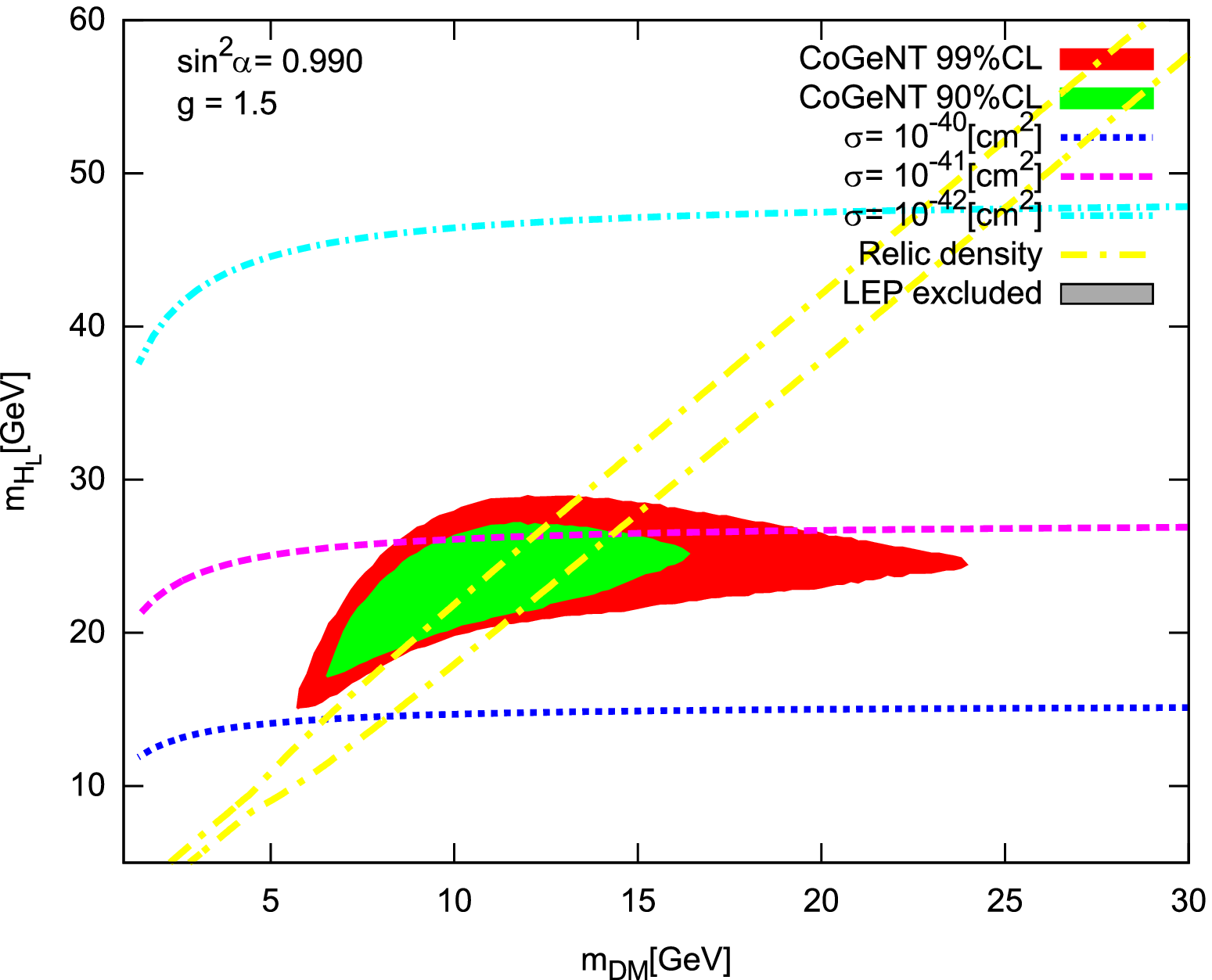}
\caption{The yellow lines imply that the correct DM relic density is
 obtained. The red and green regions show the favored parameter space by
 CoGeNT. The blue, violet and light blue lines are contours of the SI
 elastic cross section.}
\label{fig:wmap-dd}
\end{center}
\end{figure}

The contours of the SI elastic cross section $\sigma_{\mathrm{SI}}^p$
with proton for $\sin\alpha=0.980,\:0.985,\:0.990$ and
$g=0.5,\:1.0,\:1.5$ are shown in Fig.~\ref{fig:wmap-dd} in the $m_{DM}$
- $m_{H_L}$ plane. The green and red colored regions are 90\% and 99\%
confidence levels (CL) for CoGeNT. The yellow lines show that the
annihilation cross section of DM is $\sigma
v_{\mathrm{rel}}\simeq3.0\times10^{-9}~\mathrm{GeV^{-2}}$ required to satisfy the
correct DM relic density. 
The dark colored regions are excluded by the Higgs mixing bound from LEP
experiment. There is no LEP constraint for $\sin\alpha=0.990$ as seen
from Fig.~\ref{lep}. From the figures, we can see that the DM relic
density lines and 90\% CL region of CoGeNT are consistent when
$\sin\alpha$ and $g$ are $0.980-0.990$ and $1.0-1.5$. 
In the case of $g=0.5$, we need a narrow resonance point in order to
satisfy the DM relic density since Yukawa coupling is too small. 
Instead of that, too small elastic cross section is obtained in that case. 
Therefore rather large Yukawa coupling $g$ is required to be consistent
with these two aspects of DM in the allowed Higgs mixing region by LEP.


\section{Conclusions}
The Higgs portal DM model which includes a singlet real scalar
DM $S$ is the minimal extension of the SM. 
If a few GeV DM is taken into account as CoGeNT and DAMA suggest it, 
the DM in the model can be consistent with the DM relic density and
the rather large elastic cross section favored by CoGeNT. 
However, no excess of the anti-proton in the cosmic-ray
emerges in this case since the suppression by DM mass for the source
term of the anti-proton flux from the DM annihilation is no longer
effective, and the annihilation cross section includes the dominant s-wave. 

To improve it we have considered that a next minimal DM model
which includes a new Dirac fermion into the Higgs portal minimal DM model
and have reassigned the $\mathbb{Z}_2$ parity so that the Dirac fermion
can be DM candidate. 
We found that the fermionic DM is promising in the simplest model as a
result of analyzing the anti-proton no excess and the relic density. 
Since the DM has no s-wave in the annihilation cross section and has
only p-wave, the source term of anti-proton flux from the DM
annihilation was strongly suppressed at the present Universe.

On the other hand, the SM Higgs $h$ and singlet scalar $S$ mix after the
electroweak symmetry breaking in the next minimal model. 
Thus the $H_LZZ$ coupling for the light
eigenstate of Higgs is severely constrained by the LEP data. 
The allowed parameter space of the mixing angle implies small 
interaction of Higgs $\mu_SS(H^{\dag}H)$. 
In numerical analyses, we have found that the DM relic density lines and 90\% CL
parameter region of CoGeNT can be consistent when the Higgs mixing angle
$\sin\alpha$ and the Yukawa coupling $g$ are $0.980-0.990$ and $1.0-1.5$
respectively. Thus rather large Yukawa coupling $g$ is favored. 
Although only narrow Higgs mixing angle is allowed from the LEP data, 
the DM relic density and the CoGeNT favored SI elastic cross section 
can be consistent in the next minimal Dirac DM model. 

\section*{Acknowledgments}
We thank to Dr. Yuji Kajiyama for fruitful discussion.  
H.O. thanks to Prof. Eung-Jin Chun, Dr. Priyotosh Bandyopadhyay, and
Dr. Jong-Chul Park, for useful discussion of Higgs phenomenologies.
This work is supported by Young Researcher Overseas Visits Program for
Vitalizing Brain Circulation Japanese in JSPS (T.T.). 


\begin{thebibliography}{99}
\bibitem{Begeman:1991iy}
  K.~G.~Begeman, A.~H.~Broeils and R.~H.~Sanders,
  Mon.\ Not.\ Roy.\ Astron.\ Soc.\  {\bf 249}, 523 (1991).

\bibitem{wmap}
  E.~Komatsu {\it et al.}  [WMAP Collaboration],
  Astrophys.\ J.\ Suppl.\  {\bf 192}, 18 (2011)
  [arXiv:1001.4538 [astro-ph.CO]].

\bibitem{Massey:2007wb}
  R.~Massey {\it et al.},
  Nature {\bf 445}, 286 (2007)
  [arXiv:astro-ph/0701594].

\bibitem{Aprile:2011hi} 
  E.~Aprile {\it et al.}  [XENON100 Collaboration],
  Phys.\ Rev.\ Lett.\  {\bf 107}, 131302 (2011)
  [arXiv:1104.2549 [astro-ph.CO]].

\bibitem{Ahmed:2009zw} 
  Z.~Ahmed {\it et al.}  [The CDMS-II Collaboration],
  Science {\bf 327}, 1619 (2010)
  [arXiv:0912.3592 [astro-ph.CO]].

\bibitem{Angloher:2011uu} 
  G.~Angloher, M.~Bauer, I.~Bavykina, A.~Bento, C.~Bucci, C.~Ciemniak,
	G.~Deuter and F.~von Feilitzsch {\it et al.},
  arXiv:1109.0702 [astro-ph.CO].

\bibitem{cogent}
  C.~E.~Aalseth {\it et al.}  [CoGeNT collaboration],
  Phys.\ Rev.\ Lett.\  {\bf 106}, 131301 (2011)
  [arXiv:1002.4703 [astro-ph.CO]].

\bibitem{Bernabei:2010mq}
  R.~Bernabei {\it et al.},
  Eur.\ Phys.\ J.\  C {\bf 67}, 39 (2010)
  [arXiv:1002.1028 [astro-ph.GA]].

\bibitem{TuckerSmith:2001hy} 
  D.~Tucker-Smith and N.~Weiner,
  Phys.\ Rev.\ D {\bf 64}, 043502 (2001)
  [hep-ph/0101138].

\bibitem{TuckerSmith:2004jv} 
  D.~Tucker-Smith and N.~Weiner,
  Phys.\ Rev.\ D {\bf 72}, 063509 (2005)
  [hep-ph/0402065].

\bibitem{Gustafsson:2006gr} 
  M.~Gustafsson, M.~Fairbairn and J.~Sommer-Larsen,
  Phys.\ Rev.\ D {\bf 74}, 123522 (2006)
  [astro-ph/0608634].

\bibitem{Ling:2009eh} 
  F.~S.~Ling, E.~Nezri, E.~Athanassoula and R.~Teyssier,
  JCAP {\bf 1002}, 012 (2010)
  [arXiv:0909.2028 [astro-ph.GA]].

\bibitem{Andreas:2010dz}
  S.~Andreas, C.~Arina, T.~Hambye, F.~S.~Ling and M.~H.~G.~Tytgat,
  Phys.\ Rev.\  D {\bf 82}, 043522 (2010)
  [arXiv:1003.2595 [hep-ph]].

\bibitem{Evoli:2011id}
  C.~Evoli, I.~Cholis, D.~Grasso, L.~Maccione and P.~Ullio,
  arXiv:1108.0664 [astro-ph.HE].

\bibitem{arXiv:1007.5253} 
  J.~Lavalle,
  Phys.\ Rev.\ D\ {\bf 82}, 081302  (2010)
  [arXiv:1007.5253 [astro-ph.HE]].

\bibitem{Kappl:2011jw} 
  R.~Kappl and M.~W.~Winkler,
  arXiv:1110.4376 [hep-ph].

\bibitem{Goudelis:2009zz} 
  A.~Goudelis, Y.~Mambrini and C.~Yaguna,
  JCAP {\bf 0912}, 008 (2009)
  [arXiv:0909.2799 [hep-ph]].

\bibitem{Keung:2010tu} 
  W.~-Y.~Keung, I.~Low and G.~Shaughnessy,
  Phys.\ Rev.\ D {\bf 82}, 115019 (2010)
  [arXiv:1010.1774 [hep-ph]].

\bibitem{Cerdeno:2011tf} 
  D.~G.~Cerdeno, T.~Delahaye and J.~Lavalle,
  Nucl.\ Phys.\ B {\bf 854}, 738 (2012)
  [arXiv:1108.1128 [hep-ph]].

\bibitem{lep} 
  R.~Barate {\it et al.}  [LEP Working Group for Higgs boson searches
	and ALEPH and DELPHI and L3 and OPAL Collaborations],
  Phys.\ Lett.\ B {\bf 565}, 61 (2003)
  [hep-ex/0306033].

\bibitem{Navarro:1996gj}
J.F.~Navarro, C.S.~Frenk and S.D.M.~White,
  Astrophys.\ J.\ {\bf 490} (1997) 493.

\bibitem{PDG2010}
  K.~Nakamura {\it et al.}  [Particle Data Group],
  J.\ Phys.\ G {\bf 37}, 075021 (2010).

\bibitem{cms-atlas} 
  S.~Chatrchyan {\it et al.}  [CMS Collaboration],
 arXiv:1202.1488 [hep-ex];
  [ATLAS Collaboration],
  arXiv:1202.1408 [hep-ex].


\bibitem{Adriani:2010rc}
  O.~Adriani {\it et al.}  [PAMELA Collaboration],
  Phys.\ Rev.\ Lett.\  {\bf 105}, 121101 (2010)
  [arXiv:1007.0821 [astro-ph.HE]].


\bibitem{Aguilar:2002ad}
  M.~Aguilar {\it et al.}  [AMS Collaboration],
  Phys.\ Rept.\  {\bf 366}, 331 (2002)
  [Erratum-ibid.\  {\bf 380}, 97 (2003)].

\bibitem{Asaoka:2001fv}
  Y.~Asaoka {\it et al.},
  Phys.\ Rev.\ Lett.\  {\bf 88}, 051101 (2002)
  [arXiv:astro-ph/0109007].

\bibitem{Maeno:2000qx}
  T.~Maeno {\it et al.}  [BESS Collaboration],
  Astropart.\ Phys.\  {\bf 16}, 121 (2001)
  [arXiv:astro-ph/0010381].

\bibitem{Orito:1999re}
  S.~Orito {\it et al.}  [BESS Collaboration],
  Phys.\ Rev.\ Lett.\  {\bf 84}, 1078 (2000)
  [arXiv:astro-ph/9906426].

\bibitem{Burgess:2000yq} 
  C.~P.~Burgess, M.~Pospelov and T.~ter Veldhuis,
  Nucl.\ Phys.\ B {\bf 619}, 709 (2001)
  [hep-ph/0011335].

\bibitem{Barger:2007im} 
  V.~Barger, P.~Langacker, M.~McCaskey, M.~J.~Ramsey-Musolf and
	G.~Shaughnessy,
  Phys.\ Rev.\ D {\bf 77}, 035005 (2008)
  [arXiv:0706.4311 [hep-ph]].

\bibitem{Andreas:2008xy} 
  S.~Andreas, T.~Hambye and M.~H.~G.~Tytgat,
  JCAP {\bf 0810}, 034 (2008)
  [arXiv:0808.0255 [hep-ph]].

\bibitem{He:2009yd} 
  X.~-G.~He, T.~Li, X.~-Q.~Li, J.~Tandean and H.~-C.~Tsai,
  Phys.\ Lett.\ B {\bf 688}, 332 (2010)
  [arXiv:0912.4722 [hep-ph]].

\bibitem{Baek:2011aa} 
  S.~Baek, P.~Ko and W.~-I.~Park,
  JHEP {\bf 1202}, 047 (2012)
  [arXiv:1112.1847 [hep-ph]].

\bibitem{AlbornozVasquez:2011aa} 
  D.~Albornoz Vasquez, G.~Belanger, R.~M.~Godbole and A.~Pukhov,
  arXiv:1112.2200 [hep-ph].

\bibitem{Spergel:1999mh} 
  D.~N.~Spergel and P.~J.~Steinhardt,
  Phys.\ Rev.\ Lett.\  {\bf 84}, 3760 (2000)
  [astro-ph/9909386].

\bibitem{Markevitch:2003at} 
  M.~Markevitch, A.~H.~Gonzalez, D.~Clowe, A.~Vikhlinin, L.~David,
	W.~Forman, C.~Jones and S.~Murray {\it et al.},
  Astrophys.\ J.\  {\bf 606}, 819 (2004)
  [astro-ph/0309303].

\bibitem{Taoso:2007qk} 
  M.~Taoso, G.~Bertone and A.~Masiero,
  JCAP {\bf 0803}, 022 (2008)
  [arXiv:0711.4996 [astro-ph]].

\bibitem{Loeb:2010gj} 
  A.~Loeb and N.~Weiner,
  Phys.\ Rev.\ Lett.\  {\bf 106}, 171302 (2011)
  [arXiv:1011.6374 [astro-ph.CO]].

\bibitem{Buckley:2009in} 
  M.~R.~Buckley and P.~J.~Fox,
  Phys.\ Rev.\ D {\bf 81}, 083522 (2010)
  [arXiv:0911.3898 [hep-ph]].

\bibitem{Hooper:2012ft}
  D.~Hooper,
  arXiv:1201.1303 [astro-ph.CO].


\bibitem{Ohki:2008ff}
  H.~Ohki {\it et al.},
  Phys.\ Rev.\  D {\bf 78}, 054502 (2008)
  [arXiv:0806.4744 [hep-lat]].


\bibitem{Corsetti:2000yq}
  A.~Corsetti and P.~Nath,
  Phys.\ Rev.\  D {\bf 64}, 125010 (2001)
  [arXiv:hep-ph/0003186].

\end{thebibliography}
\end{document}